\documentclass[prb,aps,twocolumn,superscriptaddress]{revtex4-2}
\usepackage{graphicx}
\usepackage{dcolumn}
\usepackage{bm}
\usepackage{lipsum}
\usepackage{ragged2e}
\usepackage{braket}
\usepackage{notes2bib}
\usepackage{bbm}
\usepackage{lipsum}
\usepackage{color}
\usepackage{xcolor}
\usepackage{amsfonts}
\usepackage{amsmath}
\usepackage{amsmath,amssymb,latexsym}
\usepackage{wasysym,stmaryrd,marvosym}
\usepackage{mathrsfs}
\usepackage{soul}
\usepackage{dsfont}
\usepackage{float}

\usepackage[mathlines]{lineno}
\usepackage[colorinlistoftodos]{todonotes}
\usepackage[colorlinks=true, allcolors=blue]{hyperref}

\newcommand{\MA}[1]{{\color{blue}{#1}}}

\begin{document}

\newcommand{\ohio}{Department of Physics and Astronomy and Nanoscale and Quantum Phenomena Institute, Ohio University, Athens, Ohio 45701}

\title{Majorana edge states in $s$-wave kagome superconductors with Rashba interaction}

\author{M. A. Mojarro}
\email{mm232521@ohio.edu}
\author{Sergio E. Ulloa}
\affiliation{\ohio}

\date{\today}

\begin{abstract}
We study kagome lattices with on-site and extended spin-singlet $s$-wave superconducting pairing and show that the inclusion of Rashba spin-orbit (RSO) interaction allows time-reversal-invariant topological superconducting states which support helical Majorana pairs at the edge.
We calculate the $\mathbb{Z}_2$ topological invariant as a function of the pairing parameters for different chemical potentials. 
The rich phase diagrams reveal topological, nodal, and trivial superconducting states depending on the system parameters. 
We also consider a $2\times2$ time-reversal symmetry-breaking chiral flux phase, which has been demonstrated to be energetically favorable in the $A$V$_3$Sb$_5$ family of superconductors. Incorporating such symmetry-breaking order in our model leads to chiral Majorana edge states defined by a Chern number. 
We show how the RSO interaction allows for topological phases with even and odd Chern numbers for different system parameters. 
This work demonstrates how a simple $s$-wave kagome superconductor with RSO interaction can support helical and chiral Majorana edge states, and motivates the search for Majorana fermions in kagome superconductors. 
\end{abstract}


\maketitle

\section{Introduction}

The exploration of Majorana physics in solid-state platforms has become a prominent field of research in the last decade, as the experimental observation of Majorana zero modes (MZM's) in topological superconductors may pave the way for error-correction-free topological quantum computation thanks to their unique non-Abelian braiding properties \cite{Nayak2008,Sarma2015,Stanescu2016,Sato2017,Aghaee2025}.
The canonical model for topological superconductivity in a two-dimensional system relies on a time-reversal symmetry ($\mathcal{T}$) breaking 
$p_x+ip_y$ superconducting pairing potential (belonging to class $D$ \cite{Schnyder2008}), 
where the system has a full bulk gap and gapless chiral Majorana states propagate at the edge, while MZM's are trapped at vortex cores \cite{Read2000,Ivanov}. 
Topological superconductivity can also be induced through proximity effects enabled by a superconductor featuring a simple spin-singlet $s$-wave pairing. 
This mechanism has been proposed in several systems including quantum anomalous Hall materials \cite{Qi20102}, the surface of topological insulators \cite{Fu2008}, and spin-orbit coupled semiconductors with induced Zeeman splitting \cite{Sau2010}.
In contrast to these class-$D$ states, in a $\mathcal{T}$-invariant topological superconductor (class $DIII$), helical Majorana states propagate along the edge of the system allowed by $p_x\pm ip_y$ paring for spin-up and spin-down electrons, respectively, and a Kramers pair of MZM's are contained at a vortex core \cite{Qi2009}. 
This regime could also be achieved in a system with strong Rashba spin-orbit (RSO) coupling by proximity with a simple $s$-wave superconductor \cite{Zhang2013}, or ``intrinsically" in $s$-wave iron-based superconductors under an applied electric field \cite{Zhang2021}.  

The possibility of intrinsic topological superconductivity is naturally interesting, and the absence of proximity mechanisms results in immediate advantages. 
Strong candidates for hosting topological superconducting phases are in the family of kagome superconductors $A$V$_3$Sb$_5$ ($A=$K, Rb, Cs) \cite{Ortiz2019,Jiang2022,Wilson2024}, as they innately host both topological bands \cite{Ortiz2020,Yong2022} and superconducting order, with experiments pointing towards an $s$-wave superconducting pairing symmetry \cite{Mu2021, Duan2021,Yin2021}.
The interplay between topology and superconductivity in this family of kagome compounds suggests the possibility of Majorana physics, as suggested by STM experiments where a zero-bias conductance peak in a vortex core resembles a MZM \cite{Liang2021}.

Several studies reveal that spin-singlet superconducting pairings are the more energetically favorable in the kagome lattice \cite{Astrid2022,Wen2022}, 
and quite robust against disorder \cite{Sofie2023}.
However, most of the exploration of topological superconductivity in the kagome lattice has been based on complex superconducting order parameters which break $\mathcal{T}$ and involve spin-triplet pairing \cite{Astrid2022,Wen2022,Ding2022,Sofie2023,Liu2024}.
Unfortunately, a non-topological spin-singlet $s$-wave pairing by itself does not provide a platform for MZM's in kagome lattices \cite{Ding2022},
although breaking $\mathcal{T}$ by applying a magnetic field results in the emergence of MZM's localized at the edge of the lattice \cite{Kheirkhah2022}. 
It has also been demonstrated that a $\mathcal{T}$-preserving topological superconducting state could be achieved in the kagome lattice by proximity to a $d$-wave superconductor and in the presence of intrinsic spin-orbit coupling \cite{Kheirkhah2022}.

\begin{figure*}
 \centering
 \includegraphics[scale=0.34]{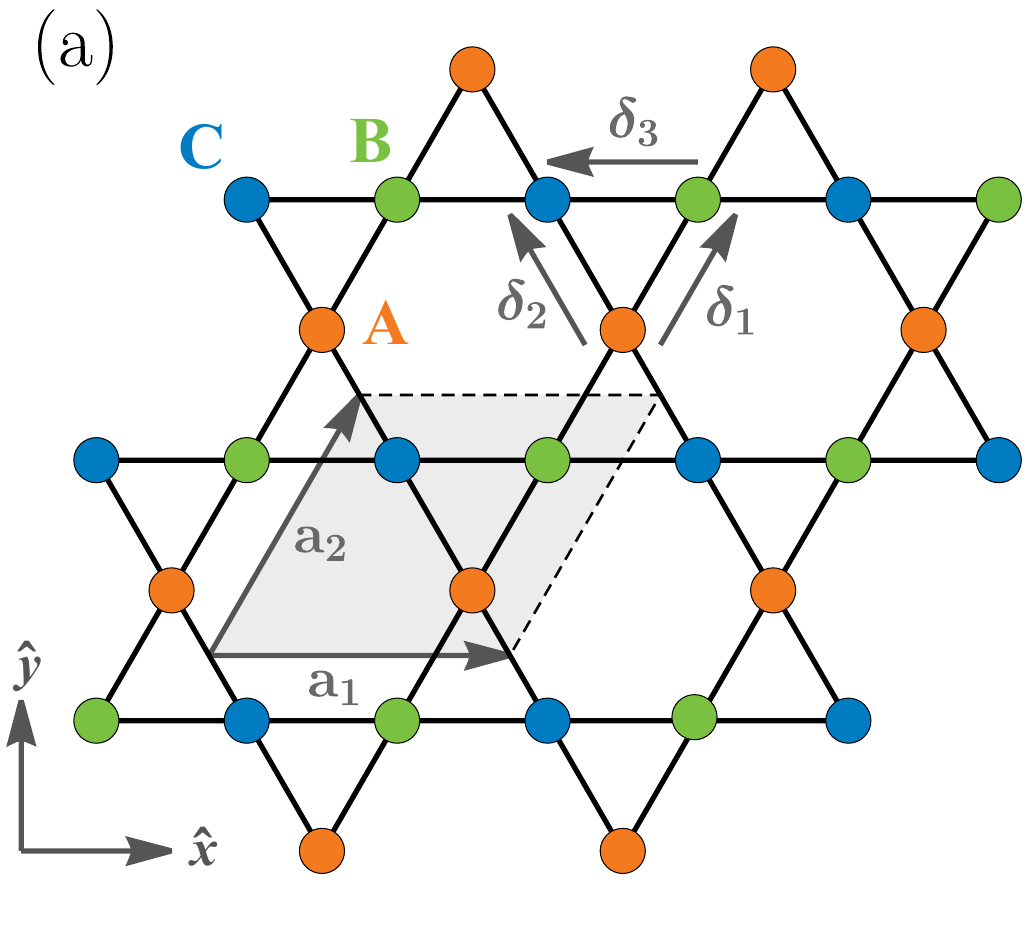}
    \includegraphics[scale=0.65]{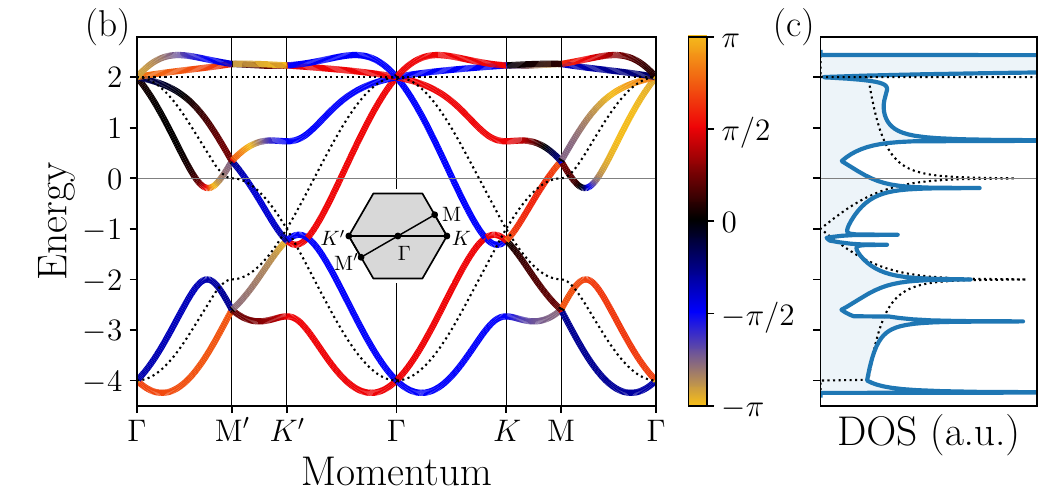}
    \caption{(a) Crystal structure of the kagome lattice, where nearest-neighbor vectors $\bm{\delta}_i$ and primitive vectors ${\bf a}_i$ are displayed. The gray region indicates the unit cell made of three sites: A (orange), B (green), and C (blue). 
    (b) Energy bands of the kagome lattice with RSO interaction ($\lambda=0.5$) along a high-symmetry path of the Brillouin zone (inset). The color bar represents the azimuthal angle of the spin orientation in momentum space for electrons in sublattice A. The corresponding density of states (DOS) is shown in (c). Dotted lines indicate the case with $\lambda=0$. 
    }
    \label{Fig1}
\end{figure*}

In this work, we explore the possibility of $\mathcal{T}$-invariant topological superconductivity in the kagome lattice with RSO coupling featuring spin-singlet $s$-wave pairing. 
We have included $s$-wave pairing between nearest-neighbor sites as it has been shown to be the leading pairing symmetry among the allowed channels in the kagome lattice \cite{Astrid2022}.
Through calculations of the $\mathbb{Z}_2$ topological invariant, we show that helical Majorana states propagate along the edge of the lattice in certain regions of parameter space, demonstrating that a simple order parameter could result in topological superconductivity in the kagome lattice.  

To explore the possibility of $\mathcal{T}$-breaking topological superconductivity in the absence of external magnetic fields, we consider a chiral flux phase (CFP) in the model \cite{Xilin2021,Denner2021}, as it has been demonstrated to lead to chiral Majorana edge states in the kagome lattice with spin-singlet pairings \cite{Zhou2023,Zeng2023}.
This phase has been suggested as the leading charge ordering instability in the $A$V$_3$Sb$_5$ family, which would help understand the observed $\mathcal{T}$-breaking of the charge order \cite{Jiang2021,Mielke2022,Guo2022}.
We find several topological phases in the kagome lattice with a CFP, each with a distinct number of chiral Majorana edge states as characterized by a Chern number. 
It is known that a $\mathcal{T}$-breaking topological superconductor characterized by an odd Chern number generally supports a MZM at a vortex core \cite{Read2000}.
When the RSO strength is distinct from zero, we find phases with odd Chern numbers in several regions of parameter space, signaling the possibility of MZM's in vortices of $s$-wave kagome superconductors which could be used for braiding operations.

The manuscript is organized as follows. In Sec. \ref{sec2} we describe the energy spectrum of the kagome lattice with RSO coupling through a tight-binding model. In Sec. \ref{sec3} we introduce our model for $\mathcal{T}$-invariant topological superconductivity in the kagome lattice. The kagome lattice with a $\mathcal{T}$-breaking CFP is presented in Sec. \ref{sec4} and the possibility for chiral Majorana edge states is discussed. Finally in Sec. \ref{sec5} we present our conclusions and discussions.

\section{kagome lattice with Rashba spin-orbit interaction}\label{sec2}


The kagome crystal structure can be derived from the line graph of the honeycomb lattice (where links are promoted to sites) \cite{Itiro1951,Daniel2018}, and consists of a triangular lattice defined by  primitive vectors:
${\bf a}_1=(2,\,0)$, ${\bf a}_2=(1,\,\sqrt{3})$, with three sites per unit cell connected by nearest-neighbor vectors: $\bm{\delta}_1=(1/2,\,\sqrt{3}/2)$, $\bm{\delta}_2=(-1/2,\,\sqrt{3}/2)$, $\bm{\delta}_3=(-1,\,0)$ (all in units of the inter-site distance),
as shown in Fig.\,\ref{Fig1}(a). The dynamics of electrons in the kagome lattice in the presence of RSO interaction is described by the following ${\cal T}$-symmetric tight-binding Hamiltonian 
\begin{eqnarray}
    \hat{\mathcal{H}}&=&\hat{\mathcal{H}}_0+\hat{\mathcal{H}}_R\,,\\
    \hat{\mathcal{H}}_0&=&-t\sum_{\braket{ij}\sigma}\hat{c}_{i\sigma}^{\dagger}\hat{c}_{j\sigma}+\text{H.c.}\,,\label{H0}\\
    \hat{\mathcal{H}}_{\text{R}}&=&i\lambda\sum_{\langle ij\rangle \sigma\sigma'}\hat{c}_{i\sigma}^\dagger[{\bm{\sigma}}_{\sigma\sigma'}\times\textbf{e}_{ij}]_z\, \hat{c}_{j\sigma'}+\text{H.c.}\,,\label{HR}
\end{eqnarray}
where $t$ is the hopping energy between two neighboring sites, $\hat{c}_{i\sigma}^\dagger$ ($\hat{c}_{i\sigma}$) is the creation (annihilation) operator of an electron at site $i$ with spin $\sigma=\,\uparrow,\,\downarrow$, $\lambda$ is the RSO coupling strength, 
$\textbf{e}_{ij}$ is a nearest-neighbor vector pointing from site $i$ to site $j$, 
and ${\bm{\sigma}}=({\sigma}_x,\,{\sigma}_y,\,{\sigma}_z)$ is a vector of Pauli matrices acting on spin space. 
Here $\braket{ij}$ indicates sum over nearest-neighbor sites and H.c. is the Hermitian conjugate. In the following we take $t=1$ as energy unit for simplicity.

After Fourier transforming the  Hamiltonian we obtain $\hat{\mathcal{H}}=\sum_{\bf k}\hat{h}_{\bf k}^\dagger H({\bf k})\hat{h}_{\bf k}$, where ${\bf k}=(k_x,\, k_y)$ is the electron wave vector and $\hat{h}_{\bf k}=(\hat{c}_{\text{A}{\bf k}\uparrow},\,\hat{c}_{\text{B}{\bf k}\uparrow},\,\hat{c}_{\text{C}{\bf k}\uparrow},\,\hat{c}_{\text{A}{\bf k}\downarrow},\,\hat{c}_{\text{B}{\bf k}\downarrow},\,\hat{c}_{\text{C}{\bf k}\downarrow})^{\text{T}}$ contains the operators representing electrons at sublattices A, B, C (see Fig.\,\ref{Fig1}(a)). 
The $6\times6$ Bloch Hamiltonian $H({\bf k})$ is
\begin{equation}\label{bloch}
  H(\textbf{k})=\begin{pmatrix}
  H_{0}(\textbf{k}) & \lambda H_{\text{R}}^{\dagger}(\textbf{k})\\
  \lambda H_{\text{R}}(\textbf{k}) & H_{0}(\textbf{k})
  \end{pmatrix},
\end{equation}
with
\begin{eqnarray}
     H_{0}({\bf k})&=&-
    \begin{pmatrix}
    0 & f_+(k_1) & f_+(k_2)\\
   f_+^*(k_1) & 0 & f_+(k_3) \\
    f_+^*(k_2)& f_+^*(k_3) & 0
    \end{pmatrix}\,,\\
    \nonumber H_{\text{R}}(\textbf{k})&=&
    \begin{pmatrix}
    0 & e^{i\pi/3}f_-(k_1) & e^{i2\pi/3}f_-(k_2)\\
    -e^{i\pi/3}f_-^*(k_1) & 0 & -f_-(k_3) \\
   -e^{i2\pi/3}f_-^*(k_2) & f_-^*(k_3) & 0
    \end{pmatrix}\,,\label{HR2}\\
\end{eqnarray}
where we have defined $f_{\pm}(x)=e^{2i x}\pm1$ and $k_i={\bf k}\cdot{\bm{\delta}}_i$. In the absence of RSO coupling ($\lambda=0$), the well-known dispersion relation of the pristine kagome lattice reads \cite{Guo2009} $\varepsilon_1({\bf k})=2$, $\varepsilon_2({\bf k})=-1+\sqrt{3+2\gamma({\bf k})}$, and $\varepsilon_3({\bf k})=-1-\sqrt{3+2\gamma({\bf k})}$, where $\gamma({\bf k})=\sum_{i=1}^3\cos 2k_i$, which consists of a flat band ($\varepsilon_1$), and two graphene-like bands ($\varepsilon_2$ and $\varepsilon_3$), as shown in Fig.\,\ref{Fig1}(b) (dotted lines).

\begin{figure}
    \centering
    \includegraphics[scale=0.4]{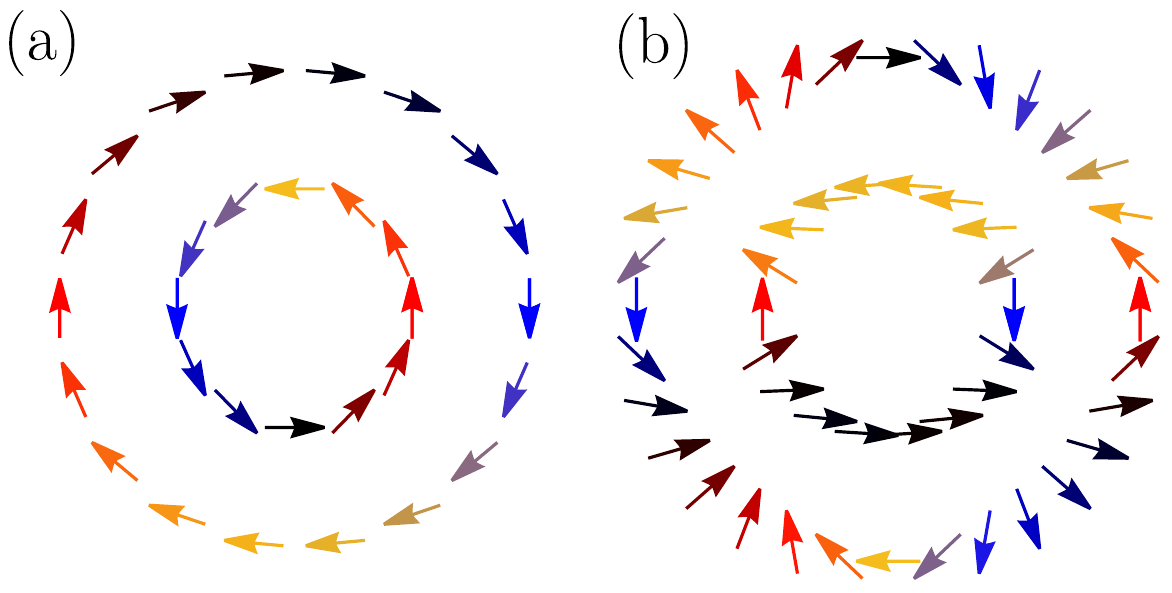}
    \caption{Schematics of spin orientation in momentum space in the vicinity of the $\Gamma$ point along Fermi contours for Fermi energies (a) $\varepsilon_F\gtrsim-4$ and (b) $\varepsilon_F\lesssim2$ (see Fig.\,\ref{Fig1}(b)). Color represents spin orientation as in Fig.\,\ref{Fig1}(b).}
    \label{spin}
\end{figure}


The RSO interaction locks the spin to the electron wave vector and breaks the spin degeneracy by coupling spin-up and spin-down states. This effect may be induced by an out-of-plane electric field which is seen as a momentum-dependent magnetic field by the electron spin \cite{Manchon2015,Bihlmayer2022}.
Strong RSO coupling due to admixture of $d$-orbitals could be realized in kagome superconductors by applying strain as inversion symmetry is broken by this distortion \cite{Tan2024}, similar to what has been seen in other materials \cite{Yao2017}. 
Fig.\,\ref{Fig1}(b) shows the energy spectrum with $\lambda=0.5$ \cite{note1},
where the color of the bands represents the spin orientation in the $k_x$\nobreakdash-$k_y$~plane of electrons in sublattice A.
The spin-degenerate energy bands when $\lambda=0$ split into two bands when RSO coupling is considered.
Such splitting leads to different families of disconnected Fermi contours. Tuning the chemical potential opens the possibility of topological superconductivity when finite electronic pairing is considered, as we will show below.


In the vicinity of the $\Gamma$ point, the lower locally-parabolic band resembles a two-dimensional electron gas (2DEG) dispersion with linear-in-momentum RSO interaction \cite{Manchon2015,Bihlmayer2022}. 
This can be seen by projecting each block in \eqref{bloch} into the eigenstate of the lower band at the $\Gamma$ point when $\lambda=0$: $\ket{\varphi_{\Gamma}}=(1,\,1,\,1)^{\text{T}}/\sqrt{3}$, 
recovering to leading order in momentum $H^{\text{low}}_{\Gamma}({\bf k})=(k^2-4)\mathbb{I}_{\sigma}+2\lambda(k_x\sigma_y-k_y\sigma_x)$, where $\mathbb{I}_{\sigma}$ is the identity matrix on spin space and $k^2=k_x^2+k_y^2$. 
The eigenstates of this Hamiltonian $\ket{{\pm{\bf k}}}$ acquire opposite chiralities $\pm$, such that the spin orientation in momentum space is $\hbar/2\braket{{\pm{\bf k}}|{\bm{\sigma}}|{\pm{\bf k}}}=\pm\hbar/2\,\hat{{\bf k}}\times\hat{{\bf z}}$, where $\hat{{\bf k}}={\bf k}/k$, as shown schematically in Fig.\,\ref{spin}(a). It can be seen that the electron spin acquires one full $2\pi$ winding when moving around a Fermi contour. 

On the other hand, the band crossing at the $\Gamma$ point but at the highest energies between the flat band and a locally parabolic band leads to a more complex spin texture when finite RSO coupling is included. 
By projecting the blocks of the Hamiltonian \eqref{bloch} into the subspace spanned by the states at ${\bf k}=0$ of the upper bands: $\ket{\phi_{\Gamma}}=(0,\,-1,\,1)^{\text{T}}/\sqrt{2}$, $\ket{\theta_{\Gamma}}=(2,\,-1,\,-1)^{\text{T}}/\sqrt{6}$, one recovers the following $4\times4$ Hamiltonian
\begin{eqnarray}
    \nonumber H^{\text{high}}_{\Gamma}({\bf k})-2\mathbb{I}_{\sigma}\mathbb{I}_{\tau}&=&-\left[\frac{k^2}{2}\mathbb{I}_{\sigma}+\lambda(k_x\sigma_y-k_y\sigma_x)\right]\mathbb{I}_{\tau}\\
    \nonumber&&-\,\left[k_xk_y\mathbb{I}_{\sigma}+\lambda(k_y\sigma_y-k_x\sigma_x)\right]\tau_x\\
    \nonumber&&-\,\left[\frac{k_x^2-k_y^2}{2}\mathbb{I}_{\sigma}+\lambda(k_y\sigma_x+k_x\sigma_y)\right]\tau_z\,,\\
\end{eqnarray}
where $\tau_x$, $\tau_y$, $\tau_z$ are Pauli matrices acting on the subspace spanned by $\ket{\phi_{\Gamma}}$ and $\ket{\theta_{\Gamma}}$, with $\mathbb{I}_{\tau}$ the identity matrix.
The first term on the right-hand side has a (doubly-degenerate) 2DEG with RSO coupling form, 
however, mixing between $\ket{\phi_{\Gamma}}$ and $\ket{\theta_{\Gamma}}$ leads to a more complex spin orientation. This is schematically shown in Fig.\,\ref{spin}(b) for energies right below the band degeneracy $\varepsilon_F\lesssim2$. For energies $\varepsilon_F\gtrsim2$, the spin texture acquires opposite orientation.  
In contrast to the 2DEG case, the eigenstates on different Fermi contours can be seen to acquire three full windings when moving around the outer Fermi contour, and only one full winding along the inner one. 
Notice, however, that as the spin orientation is a tangential field on a 2-torus (Brillouin zone), the total spin winding over the entire Brillouin zone vanishes as required by the Poincaré-Hopf index theorem \cite{fulton1997}. 
This spin texture in the vicinity of $\Gamma$ mimics the spin winding of a system with cubic RSO coupling, which could lead to topological $f$-wave superconductivity when coupled to an $s$-wave superconductor \cite{Mao2011,Sheng2024}.

It is worth mentioning that similar to the graphene honeycomb lattice, each Dirac point splits into two points, one of them remaining at the corner of the Brillouin zone, while the other moves towards $\Gamma$ respecting $C_6$ symmetry, and leading to a trigonal warping dispersion in the vicinity of the $K,\,K'$ points \cite{Zarea2009,Rakyta2010}. 
However, the emergent Dirac points in the kagome lattice are not degenerate in energy with the original ones at $K,\,K'$.

\section{Model for $\mathcal{T}$-invariant topological superconductivity}\label{sec3}

\begin{figure*}
 \centering
    \includegraphics[scale=0.82]{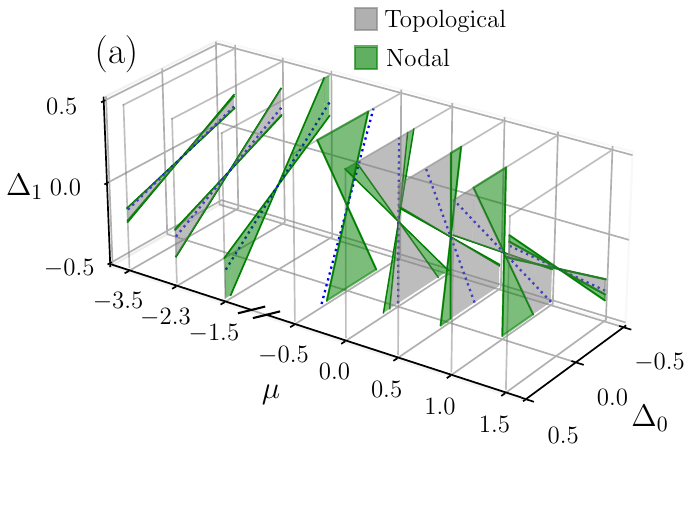}
    \includegraphics[scale=0.58]{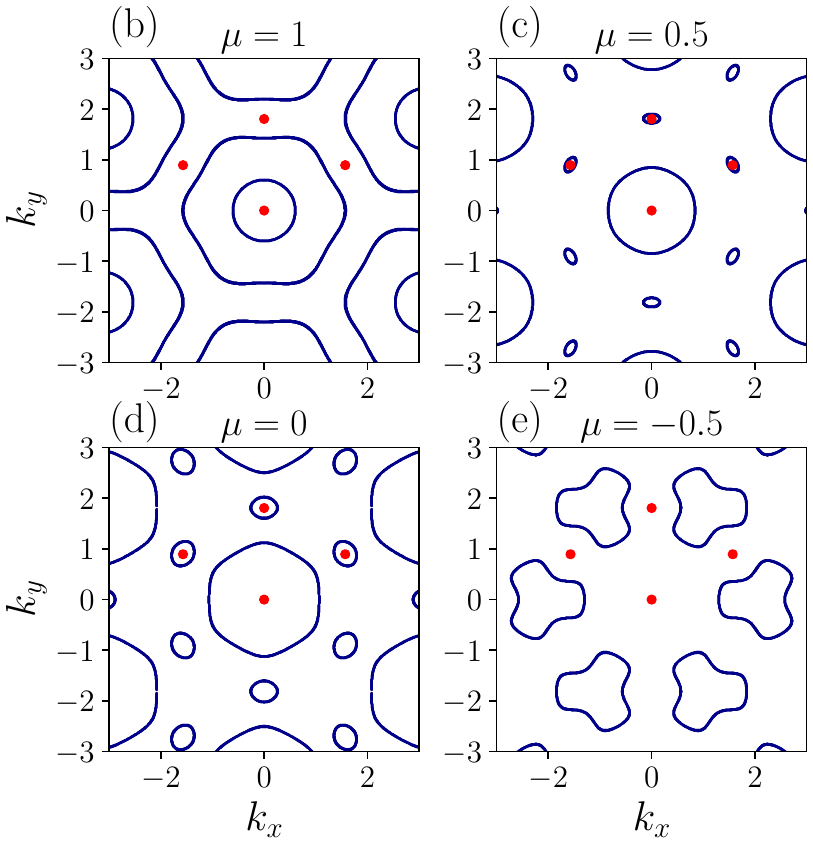}
    \caption{(a) Phase diagrams in the $(\Delta_0,\,\Delta_1)$ space for several chemical potentials. Gray and green represent topological and nodal superconducting phases, respectively. Otherwise the system is a trivial superconductor. The blue dotted lines $\Delta_0/\Delta_1=\mu$ indicate the nodal phase in the absence of RSO coupling. Notice that for $-2\lesssim\mu\lesssim-0.2$, the system is either a nodal or trivial gapped superconductor. (b), (c), (d), and (e) show the Fermi contours for $\mu=1,\,0.5,\,0,\,-0.5$ in (a), respectively. Time-reversal invariant momenta are marked as red dots.
    For $\mu=1$ the Fermi contour is made of two Fermi pockets enclosing the $\Gamma$ point.
     At $\mu=0$ and $\mu=0.5$, a single Fermi pocket encloses the $\Gamma$ point, and Fermi pockets enclose the non-equivalent M points. The pairing amplitude values at each $\mu$ depend on ($\Delta_0,\,\Delta_1$), see Fig.\,\ref{Edge}.}
    \label{PhaseD1}
\end{figure*}

To investigate the presence of helical Majorana edge states in the kagome lattice with RSO coupling we consider different pairing interactions via the model Hamiltonian
\begin{eqnarray}\label{H1}
    \hat{\mathcal{H}}&=&\hat{\mathcal{H}}_0+\hat{\mathcal{H}}_{\text{R}}+\hat{\mathcal{H}}_{\text{SC}}\,,\\
    \hat{\mathcal{H}}_{\text{SC}}&=&\Delta_0\sum_{i}\hat{c}^{\dagger}_{i\uparrow}\hat{c}^{\dagger}_{i\downarrow}+\Delta_1\sum_{\braket{ij}}\hat{c}^{\dagger}_{i\uparrow}\hat{c}^{\dagger}_{j\downarrow}+\text{H.c.}\,,
\end{eqnarray}
where $\hat{\mathcal{H}}_0$ and $\hat{\mathcal{H}}_{\text{R}}$ are given in Eqs.\,\eqref{H0} and \eqref{HR}, respectively. The $\Delta_0$ and $\Delta_1$ terms describe on-site and nearest-neighbor $s$-wave superconducting pairings, respectively \cite{Zhang2013}. 
In momentum space, the Hamiltonian in Eq. \eqref{H1} acquires a Bogoliubov-de Gennes (BdG) form as follows
\begin{equation}\label{BdG}
    H_{\text{BdG}}({\bf k})=\begin{pmatrix}
        H({\bf k}) - \mu & i\sigma_y\Delta({\bf k}) \\
        -i\sigma_y\Delta({\bf k})  &-H^{\text{T}}(-{\bf k})+\mu
    \end{pmatrix}\,,
\end{equation}
where 
\begin{equation}
    \Delta({\bf k})=   \begin{pmatrix}
    \Delta_0 & \Delta_1 f_+(k_1) & \Delta_1f_+(k_2)\\
    \Delta_1f^*_+(k_1) & \Delta_0 & \Delta_1f_+(k_3) \\
     \Delta_1f^*_+(k_2)& \Delta_1f^*_+(k_3) & \Delta_0
    \end{pmatrix}\,,
\end{equation}
and $H({\bf k})$ is given in Eq.\,\eqref{bloch}. We have also included the chemical potential explicitly to control electron filling. The $12\times12$ BdG Hamiltonian is written in the new basis $\hat{\Psi}_{\bf k}=(\hat{c}_{{\bf k}\uparrow},\,\hat{c}_{{\bf k}\downarrow},\,\hat{c}^\dagger_{-{\bf k}\uparrow},\hat{c}^\dagger_{-{\bf k}\downarrow})^{\text{T}}$, 
where $\hat{c}_{{\bf k}\sigma}$ ($\hat{c}_{{\bf k}\sigma}^{\dagger}$) contains annihilation (creation) operators associated to sublattices A, B, C.
Notice how the pair potential $i\sigma_y\Delta({\bf k})$ is antisymmetric in spin space and $\Delta({\bf k})=\Delta^{\text{T}}(-{\bf k})$, as required by a spin-singlet pairing potential. 
In the kagome lattice, superconducting pairings transform according to the irreducible representation of the point group $C_{6v}$, namely the spin singlets $A_{1g}$, $A_{2g}$, $E_{2g}$, and the spin triplets $B_{1u}$, $B_{2u}$, $E_{1u}$, where $g$ ($u$) labels even-parity (odd-parity) representations \cite{Wen2022,Sofie2023,Liu2024}.
In our case, $\Delta({\bf k})$ respects all the symmetries of the normal-state Hamiltonian, thus belonging to the $A_{1g}$ representation. 


When $\lambda=0$, diagonalization of the BdG Hamiltonian \eqref{BdG} leads to spin-degenerate quasiparticle bands $\xi_i({\bf k})=\pm\sqrt{(\varepsilon_i({\bf k})-\mu)^2+(\Delta_0-\Delta_1\varepsilon_i({\bf k}))^2}$, where each $\varepsilon_i({\bf k})$ ($i=1,2,3$) are the eigenvalues of $H_0({\bf k})$ given below Eq.\,\eqref{HR2}. 
When $\Delta_1=0$, the Bogoliubov spectrum acquires a superconducting gap $2\Delta_0$, 
and for $\Delta_1\neq0$  the system becomes a nodal superconductor whenever $\Delta_0/\Delta_1=\mu$, otherwise the system is gapped. 
When $\lambda\neq0$, as discussed before, breaking of spin degeneracy leads to splitting of the energy bands and associated disconnected Fermi contours, opening the possibility of $\mathcal{T}$-invariant topological superconducting phases as $\Delta({\bf k})$ is momentum dependent \cite{Zhang2013}.

\begin{figure}
    \centering
    \includegraphics[scale=0.195]{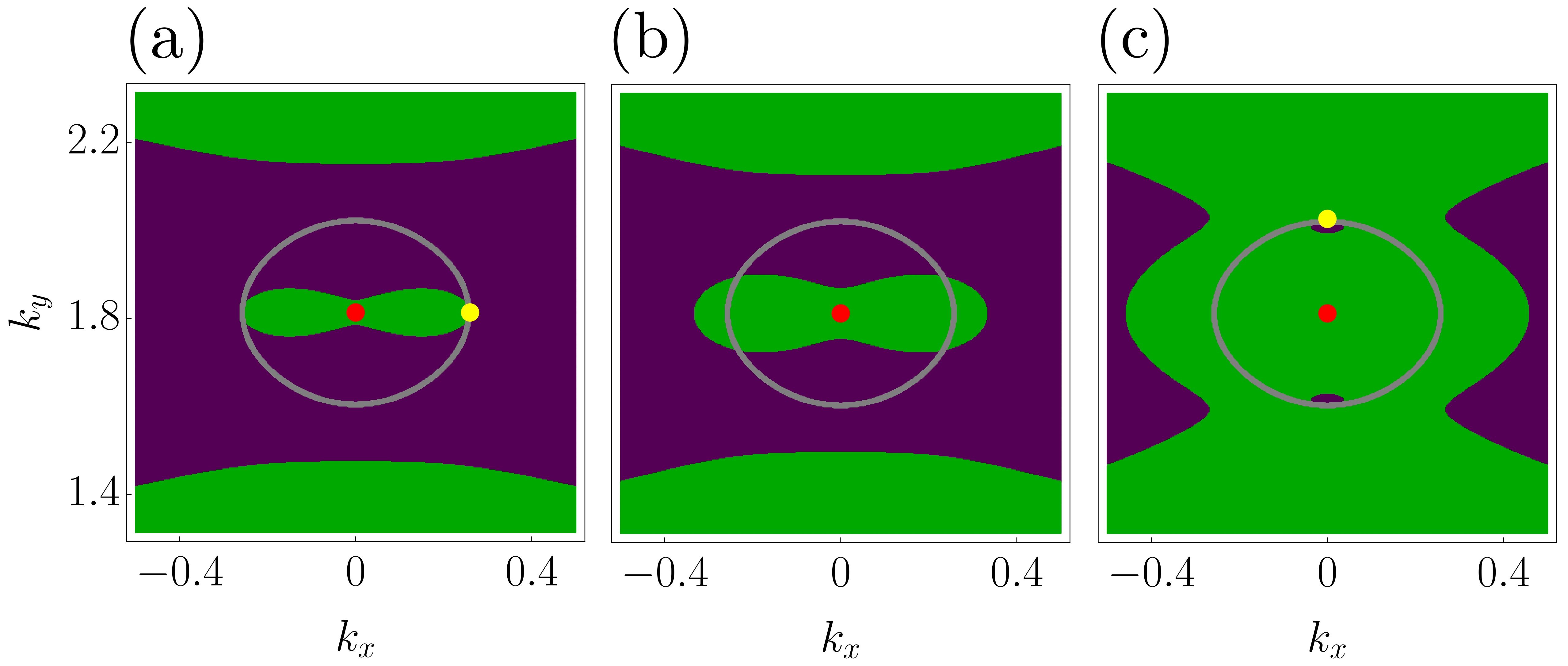}
    \caption{Evolution of sign of paring amplitude $\delta_i$ in the vicinity of the $\Gamma_3$ point (red dot) surrounded by a Fermi contour (gray line) for $\mu=0$, taking $\Delta_1=0.5$  as $\Delta_0$ increases from (a) to (c). Green and purple regions indicate positive
and negative values of the pairing.
In (a) and (c) we take $\Delta_0=\Delta_1(\mu-\lambda\epsilon_{\text{R}}({\bf k}_F^1))$ and $\Delta_0=\Delta_1(\mu-\lambda\epsilon_{\text{R}}({\bf k}_F^2))$, respectively, where ${\bf k}_F^1$ and ${\bf k}_F^2$ are indicated by yellow dots. In (b) we take an intermediate $\Delta_0$. In these three cases, the system remains nodal, as discussed in the text.}
    \label{Sgnd}
\end{figure}

In a $\mathcal{T}$-invariant superconductor, the topological characteristics depend on the properties of the Fermi surfaces surrounding $\mathcal{T}$-invariant points $\Gamma_i$ in the Brillouin zone \cite{Qi2010}, 
which satisfy $-\Gamma_i={\bf G} + \Gamma_i$, where ${\bf G}$ is a reciprocal basis vector, and $H({\Gamma}_i)=\hat{\mathcal{T}}H(\Gamma_i)\hat{\mathcal{T}}^{-1}$, 
with $\hat{\mathcal{T}}=-i\sigma_y\hat{\mathcal{K}}$ the time-reversal operator, and $\hat{\mathcal{K}}$ the complex conjugation operator. 
In two-dimensional systems, there are exactly four $\mathcal{T}$-invariant points. 
In the kagome Brillouin zone, they are located at $\Gamma_1=(0,\,0)$, and the M points $\Gamma_2=(\pi,\,\pi/\sqrt{3})/2$, $\Gamma_3=(0,\,\pi/\sqrt{3})$, and $\Gamma_4=(-\pi,\,\pi/\sqrt{3})/2$.
A $\mathcal{T}$-invariant superconductor is classified by a $\mathbb{Z}_2$ topological number $\nu$ determined by  $(-1)^{\nu}=\Pi_{j}[\text{sgn}(\delta_j)]^m$ \cite{Qi2010},
where $m$ is the number of $\mathcal{T}$-invariant points enclosed by the $j$th Fermi contour, and the pairing amplitude matrix element $\delta_j=\braket{\psi_{n{\bf k}_F}|\hat{\mathcal{T}}(i\sigma_y\Delta({\bf k}_F))^{\dagger}|\psi_{n{\bf k}_F}}$, where $\ket{\psi_{n{\bf k}_F}}$ is the eigenstate of \eqref{bloch} with eigenvalue $E_{n}({\bf k}_F)=\mu$, for some momenta ${\bf k}_F$ defining the $j$th Fermi contour. 
A non-trivial topological state is defined for $\nu=1$ and realized when the pairing is negative in an odd number of Fermi lines, each enclosing a $\mathcal{T}$-invariant point; 
otherwise the system is in a trivial phase ($\nu=0$).
In our case, $\delta_j=-\Delta_0+\Delta_1\braket{\psi_{n{\bf k}_F}|\mathbb{I}_{\sigma}H_0({\bf k}_F)|\psi_{n{\bf k}_F}}=-\Delta_0+\Delta_1(\mu-\lambda\epsilon_{\text{R}}({\bf k}_F))$, where $\epsilon_{\text{R}}({\bf k})=\braket{\psi_{n{\bf k}}|\tilde{H}_{\text{R}}({\bf k})|\psi_{n{\bf k}}}$ with $\tilde{H}_{\text{R}}({\bf k})=(\sigma_x+i\sigma_y)H_{\text{R}}^\dagger({\bf k})/2+(\sigma_x-i\sigma_y)H_{\text{R}}({\bf k})/2$, and $H_{\text{R}}({\bf k})$ given in \eqref{HR2}.

Three different families of Fermi contours are found as $\mu$ is tuned.
For $-2\lesssim\mu\lesssim-0.2$, the Fermi contours enclose no time-reversal invariant points, preventing the possibility of topological superconducting phases (see Fig.\,\ref{PhaseD1}(e)).
For other values of $\mu$, Fermi contours surround one $\mathcal{T}$-invariant point each in the Brillouin zone.
For $0.75\lesssim\mu\lesssim2.4$ and $-4.2\lesssim\mu\lesssim-2.4$, two Fermi contours enclose the $\Gamma_1$ point as shown in Fig.\,\ref{PhaseD1}(b) taking $\mu=1$. 
On the other hand, for $-0.2\lesssim\mu\lesssim0.75$ and $-2.4\lesssim\mu\lesssim-2$, only one Fermi contour encloses the $\Gamma_1$ point, while three Fermi contours enclose the points $\Gamma_2,\Gamma_3,\Gamma_4$ (see Figs.\,\ref{PhaseD1}(c)-(d)). However, as $C_{6v}$ symmetry is preserved, those three Fermi contours are all equivalent. 
Therefore, for any $\mu\lesssim-2$ and $\mu\gtrsim-0.2$, only two non-equivalent Fermi contours (where the pairing is $\delta_1$ and $\delta_2$)  determine the topology of the system, so that for parameters where $\delta_1\delta_2<0\,(>0)$, the system is a topological (trivial) superconductor.
Phase diagrams on the ($\Delta_0$,\,$\Delta_1$) space are shown in Fig.\,\ref{PhaseD1}(a)  for several chemical potentials. 
We find three distinct phases as parameters vary, namely a topological superconducting phase (gray), a nodal phase (green), and a gapped trivial superconductor otherwise.
Topological phases exist only whenever $\Delta_1\neq0$.
In the nodal phase, the pairing $\delta_i$ changes sign when moving along a Fermi contour, so that this phase is defined whenever $\Delta_1=\Delta_0/(\mu-\lambda\epsilon_{\text{R}}({\bf k}_F))$. 
We have found that there are two wave vectors ${\bf k}^1_F$ and ${\bf k}^2_F$ on each Fermi contour such that the system remains nodal when $\Delta_0/(\mu-\lambda\epsilon_{\text{R}}({\bf k}^1_F))\leqslant\Delta_1\leqslant\Delta_0/(\mu-\lambda\epsilon_{\text{R}}({\bf k}^2_F))$ (see Fig.\,\ref{Sgnd}). 

\begin{figure}
    \centering
\includegraphics[scale=0.43]{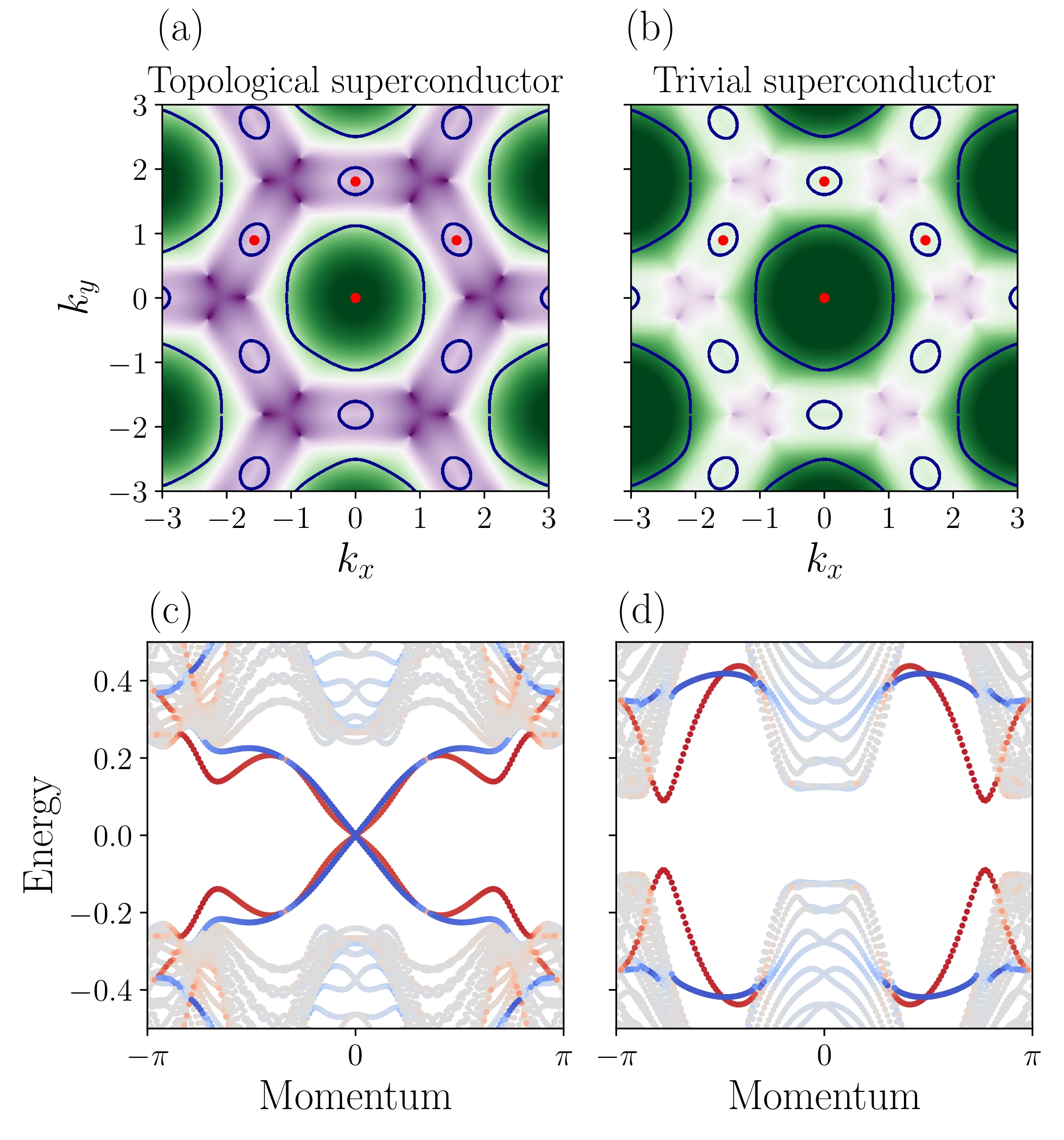}
    \caption{Map of pairing amplitude $\braket{\psi_{n{\bf k}}|\hat{\mathcal{T}}(i\sigma_y\Delta({\bf k}))^{\dagger}|\psi_{n{\bf k}}}$ for the band $n$ crossing $\mu=0$ in (a) the topological phase  $(\Delta_0,\,\Delta_1)=(0.15,\,0.5)$, and (b) the trivial phase $(\Delta_0,\,\Delta_1)=(-0.25,\,0.5)$.
    Green and purple denote positive and negative values of the pairing, respectively.   
    In the topological superconducting state (a), the pairing has opposite signs on two non-equivalent Fermi contours (blue lines) enclosing time-reversal invariant points  (red dots). In contrast, in the topologically trivial phase in (b), the pairing amplitude has positive sign on both non-equivalent Fermi contours. (c), (d) shows the energy spectrum of an infinite strip of kagome lattice with 30 unit cells in the finite direction. 
    A pair of counterpropagating Majorana edge states appear at the top (blue) and bottom (red) edges of the strip in the bulk topological superconducting gap, as shown in (c). In (d) the system is the trivial superconductor in (b) and remains gapped.}
    \label{Edge}
\end{figure}

Figs.\,\ref{Edge}(a)-(b) illustrate how the $\mathbb{Z}_2$ topological invariant is obtained by looking at the sign of the pairing along the Fermi contours. 
Fig.\,\ref{Edge}(a) displays a color map of the pairing amplitude $\braket{\psi_{n{\bf k}}|\hat{\mathcal{T}}(i\sigma_y\Delta({\bf k}))^{\dagger}|\psi_{n{\bf k}}}$ in momentum space in the topological phase for the band $n$ crossing $\mu=0$, with $(\Delta_0,\,\Delta_1)=(0.15,\,0.5)$.
We observe how the pairing acquires different signs on each of the two non-equivalent Fermi lines enclosing a $\mathcal{T}$-invariant point, leading to a topological phase as $\delta_1\delta_2<0$.
In contrast, Fig.\,\ref{Edge}(b) shows the case of a trivial superconducting phase with $(\Delta_0,\,\Delta_1)=(-0.25,\,0.5)$ and $\mu=0$, where the pairing has a positive sign on both non-equivalent Fermi contours, leading to $\delta_1\delta_2>0$.

By the bulk-boundary correspondence, a gapped system characterized by a non-trivial topological invariant will host topologically protected conducting states at its boundary.
In a $\mathcal{T}$-invariant superconductor, these states manifest as helical Majorana Kramers pairs traveling in opposite directions and with opposite spin along the edges of the sample. 
Fig.\,\ref{Edge}(c) shows the energy spectrum of an infinite strip of kagome lattice in the topological superconducting phase of \ref{Edge}(a). We observe that while the bulk remains gapped, helical gapless Majorana edge states propagate at the top and bottom of the strip, in contrast to the trivial phase shown in Fig.\,\ref{Edge}(d).

\begin{figure}
    \centering
\includegraphics[scale=0.37]{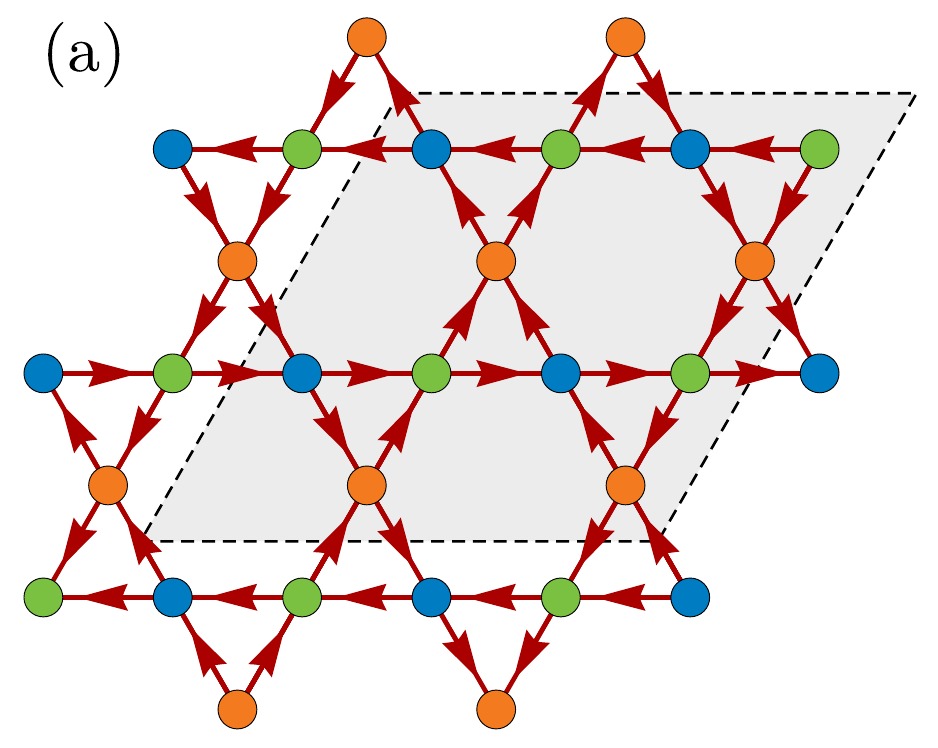} \\
 \includegraphics[scale=0.08]{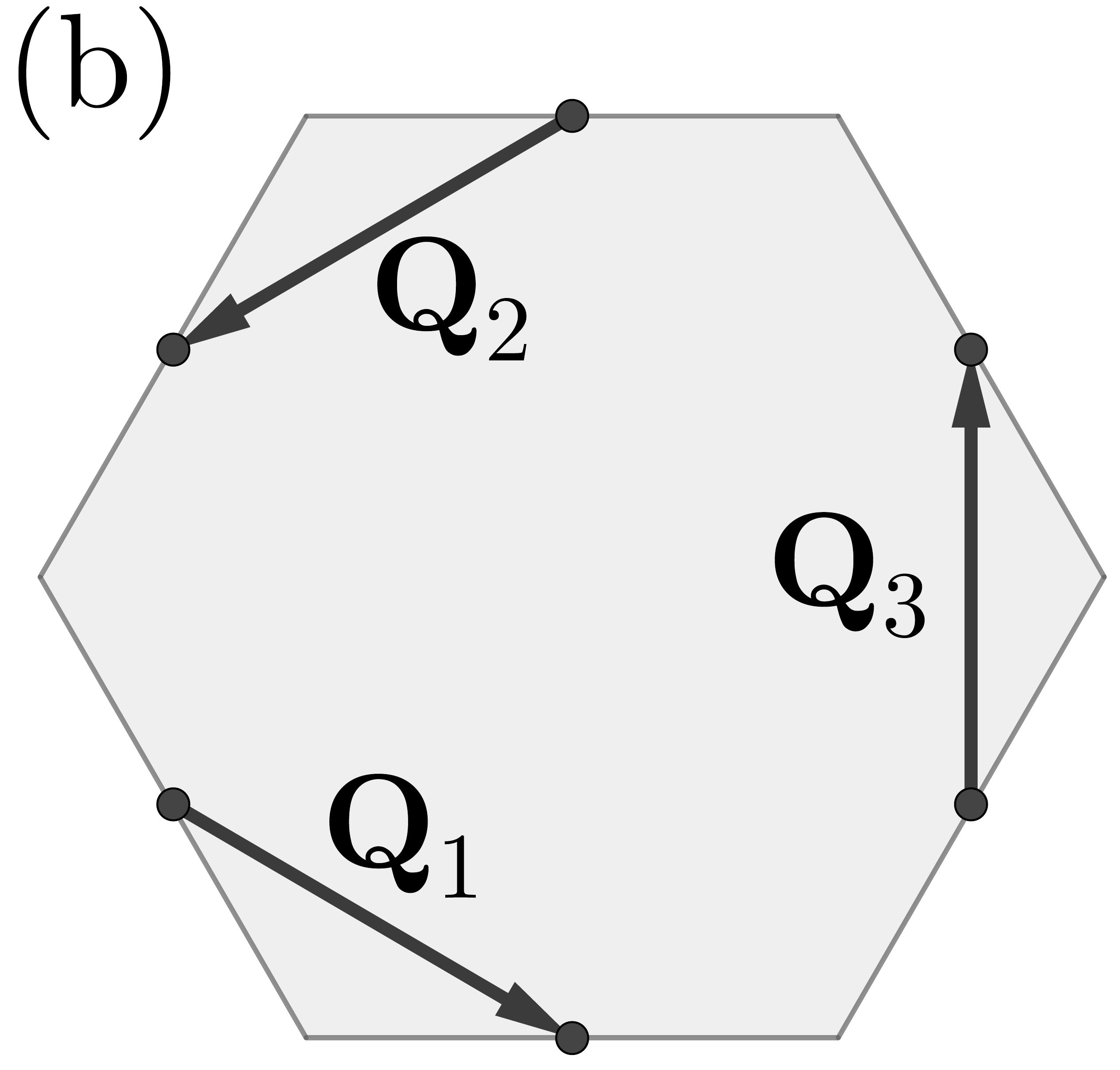} \includegraphics[scale=0.5]{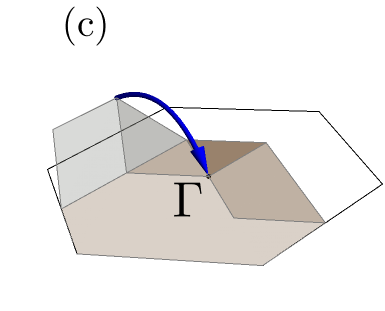} 
    \caption{(a) A $2\times2$ CFP in the kagome lattice enlarges the unit cell by 4 (shaded region) and stems from scattering between M points mediated by $\mathbf{Q}_i$ vectors as shown in (b). The direction of the red arrows indicates the flux phase. 
    This modulation leads to a Brillouin zone folding as shown in (c).}
    \label{FigCFP}
\end{figure}


\section{chiral flux phase and chiral Majorana edge states}\label{sec4}

As MZM's at vortex cores come in pairs in helical topological superconductors, we now explore the possibility of chiral superconductivity in the kagome lattice. In this phase an odd number of chiral Majorana edge states could be realized, as well as single MZM's at vortex cores. 
To this end, we introduce a $\mathcal{T}$-breaking chiral flux phase (CFP) in the model.
This phase has been identified as the leading instability in the vanadium superconductors and exhibits a $2\times2$ charge ordering \cite{Xilin2021,Denner2021,Jiang2021,Mielke2022}, see Fig.\,\ref{FigCFP}(a). 
Thus, we add an additional term to our model Hamiltonian to account for this complex bond density wave
\begin{eqnarray}\label{FullH}
    \hat{\mathcal{H}}&=&\hat{\mathcal{H}}_0+\hat{\mathcal{H}}_{\text{R}}+\hat{\mathcal{H}}_{\text{SC}}+\hat{\mathcal{H}}_{\text{CFP}}\,,\label{TotalH}\\ \label{HCFP}
    \hat{\mathcal{H}}_{\text{CFP}}&=&i\sum_{\braket{ij}\sigma}\lambda_{ij}\,\hat{c}^{\dagger}_{i\sigma}\hat{c}_{j\sigma}+\text{H.c.}\,,
\end{eqnarray}
where the bond order parameter $\lambda_{ij}$ takes the form \cite{Xilin2021}
\begin{equation}\label{LCFP}
    \lambda_{ij} = \pm\lambda_{\text{CFP}}\left\{ \begin{array}{lr} \cos({\bf Q}_1\cdot\text{\bf r}) & :i=\text{A},\,j=\text{B}\,,\\ \cos({\bf Q}_2\cdot\text{\bf r}) & :i=\text{A},\,j=\text{C}\,,\\
    \cos({\bf Q}_3\cdot\text{\bf r}) & :i=\text{B},\,j=\text{C}\,,\end{array} \right. 
\end{equation}
where the $+\,(-)$ sign corresponds to intra(inter)-cell components, with $\bf r$ the coordinate of the cell, and ${\bf Q}_1=(\pi,\,-\pi/\sqrt{3})/2$, ${\bf Q}_2=(-\pi,\,-\pi/\sqrt{3})/2$, ${\bf Q}_3=(0,\,\pi/\sqrt{3})$ are scattering wave vectors between M points (see Fig.\,\ref{FigCFP}(b)). 
The CFP strength $\lambda_{\text{CFP}}$ is taken as  $\lambda_{\text{CFP}}=0.25$ in the following. 
The remaining terms in Eq.\,\eqref{FullH} are defined as before. 
This $2\times2$ complex bond density wave enlarges the unit cell, while the Brillouin zone folds accordingly, mapping the M points back to $\Gamma$ as illustrated in Fig.\,\ref{FigCFP}(c).

\begin{figure}
    \centering
    \includegraphics[scale=0.438]{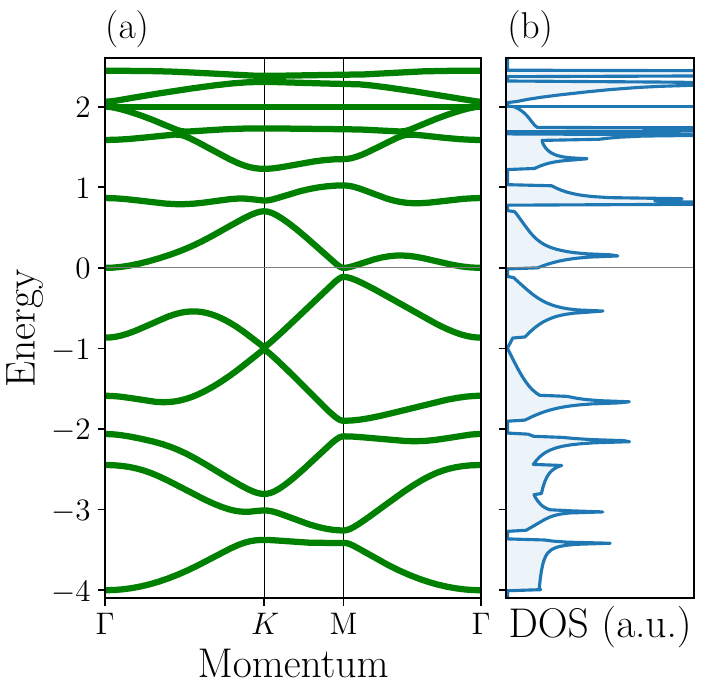}
    \includegraphics[scale=0.438]{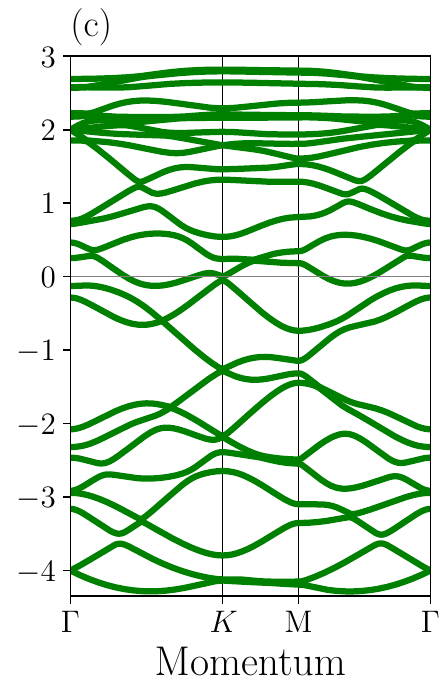}
    \caption{(a) Single-particle energy spectrum (b) and density of states of the kagome lattice with a CFP along a high symmetry path of the folded Brillouin zone in the absence of RSO coupling. (c) Energy bands with finite RSO coupling.}
    \label{CFPbands}
\end{figure}

The CFP term in the Hamiltonian (omitting spin) is given in momentum space as (see Appendix \ref{Append-A})
\begin{widetext} 
\begin{equation}
    \hat{\mathcal{H}}_{\text{CFP}}=i\lambda_{\text{CFP}}\sum_{{\bf k}}f_-(k_1)\,\hat{c}^\dagger_{\text{A}{\bf k}}\hat{c}_{\text{B}{\bf k}+{\bf Q}_1}+f_-(k_2)\,\hat{c}^\dagger_{\text{A}{\bf k}}\hat{c}_{\text{C}{\bf k}+{\bf Q}_2}-f_-(k_3)\,\hat{c}^\dagger_{\text{B}{\bf k}}\hat{c}_{\text{C}{\bf k}+{\bf Q}_3}+\text{H.c.}\,.
\end{equation}
\end{widetext} 
The periodicity of the associated Bloch Hamiltonian is determined by the original reciprocal basis vectors, 
and in order to restrict its periodicity to the superlattice Brillouin zone, we write it in the basis 
\begin{widetext} 
\begin{equation}\label{BasisF}
    \hat{h}_{\bf k}=(\hat{c}_{\text{A}{\bf k}},\,\hat{c}_{\text{B}{\bf k}},\,\hat{c}_{\text{C}{\bf k}},\,\hat{c}_{\text{A}{\bf k}+{\bf Q}_1},\,\hat{c}_{\text{B}{\bf k}+{\bf Q}_1},\,\hat{c}_{\text{C}{\bf k}+{\bf Q}_1},\,\hat{c}_{\text{A}{\bf k}+{\bf Q}_2},\,\hat{c}_{\text{B}{\bf k}+{\bf Q}_2},\,\hat{c}_{\text{C}{\bf k}+{\bf Q}_2},\,\hat{c}_{\text{A}{\bf k}+{\bf Q}_3},\,\hat{c}_{\text{B}{\bf k}+{\bf Q}_3},\,\hat{c}_{\text{C}{\bf k}+{\bf Q}_3})^{\text{T}}\,,
\end{equation}
\end{widetext} 
such that $\hat{\mathcal{H}}_{\text{CFP}}=\sum_{\bf k}\hat{h}_{\bf k}^\dagger H_{\text{CFP}}({\bf k})\hat{h}_{\bf k}$, where the $12\times12$ CFP Bloch Hamiltonian reads

\begin{eqnarray}
    \nonumber H_{\text{CFP}}({\bf k})&=&
    \begin{pmatrix}
        0 & \mathcal{V}_1({\bf 0},\,{\bf Q}_1) & \mathcal{V}_2({\bf 0},\,{\bf Q}_2) & \mathcal{V}_3({\bf 0},\,{\bf Q}_3)\\
        0 & 0  & \mathcal{V}_3({\bf Q}_1,\,{\bf Q}_2) & \mathcal{V}_2({\bf Q}_1,\,{\bf Q}_3)\\
         0 & 0 & 0 & \mathcal{V}_1({\bf Q}_2,\,{\bf Q}_3)\\
       0 & 0 & 0 & 0
    \end{pmatrix}\,,\\
    &&+\,\text{H.c.}
\end{eqnarray}
with
\begin{equation}
    \mathcal{V}_1({\bf u},\,{\bf v})=i\lambda_{\text{CFP}}\begin{pmatrix}
        0 & f_-(k_1+{\bf u}\cdot\bm{\delta}_1) & 0 \\
       -f_-^*(k_1+{\bf v}\cdot\bm{\delta}_1) &0 &0\\
       0 & 0 & 0
    \end{pmatrix}\,,
\end{equation}
\begin{equation}
    \mathcal{V}_2({\bf u},\,{\bf v})=i\lambda_{\text{CFP}}\begin{pmatrix}
        0 & 0 & f_-(k_2+{\bf u}\cdot\bm{\delta}_2) \\
       0 &0 &0\\
       - f_-^*(k_2+{\bf v}\cdot\bm{\delta}_2)& 0 & 0
    \end{pmatrix}\,,
\end{equation}
\begin{equation}
    \mathcal{V}_3({\bf u},\,{\bf v})=i\lambda_{\text{CFP}}\begin{pmatrix}
        0 & 0 & 0 \\
       0 &0 & - f_-(k_3+{\bf u}\cdot\bm{\delta}_3)\\
       0&  f_-^*(k_3+{\bf v}\cdot\bm{\delta}_3) & 0
    \end{pmatrix}\,.
\end{equation}
The total Hamiltonian \eqref{TotalH} is similarly written in the basis \eqref{BasisF} and treated accordingly when spin and superconducting pairing are considered. 
In the absence of RSO coupling and superconducting pairing, the energy bands and density of states of the kagome lattice with a CFP are shown in Fig.\,\ref{CFPbands}.

As time-reversal symmetry is broken by the CFP, a Chern number $C$ for occupied bands, when the Fermi energy lies in a gap, is related to the anomalous Hall conductance as $\sigma_{H}=Ce^2/h$, where $e$ is the electron charge and $h$ the Planck constant. The Chern number $C^{(n)}$ for an isolated band $n$ is defined as 
\begin{equation}
    C^{(n)}=\frac{1}{2\pi}\int_{\text{BZ}}d{\bf k}\,\Omega_{xy}^{(n)}({\bf k})\,,
\end{equation}   
where $\Omega_{xy}^{(n)}({\bf k})=\partial_{k_x}\mathcal{A}^{(n)}_y({\bf k})-\partial_{k_y}\mathcal{A}^{(n)}_x({\bf k})$ is the Berry curvature and 
$\mathcal{A}_i^{(n)}({\bf k})=i\braket{\psi_{n\bf k}|\partial_{k_i}|\psi_{n\bf k}}$ is the Berry connection. 
Therefore, the total Chern number is determined by $C=\sum_{\varepsilon_n<\varepsilon_F}C^{(n)}$.

\begin{figure}
    \centering
    \includegraphics[scale=0.4]{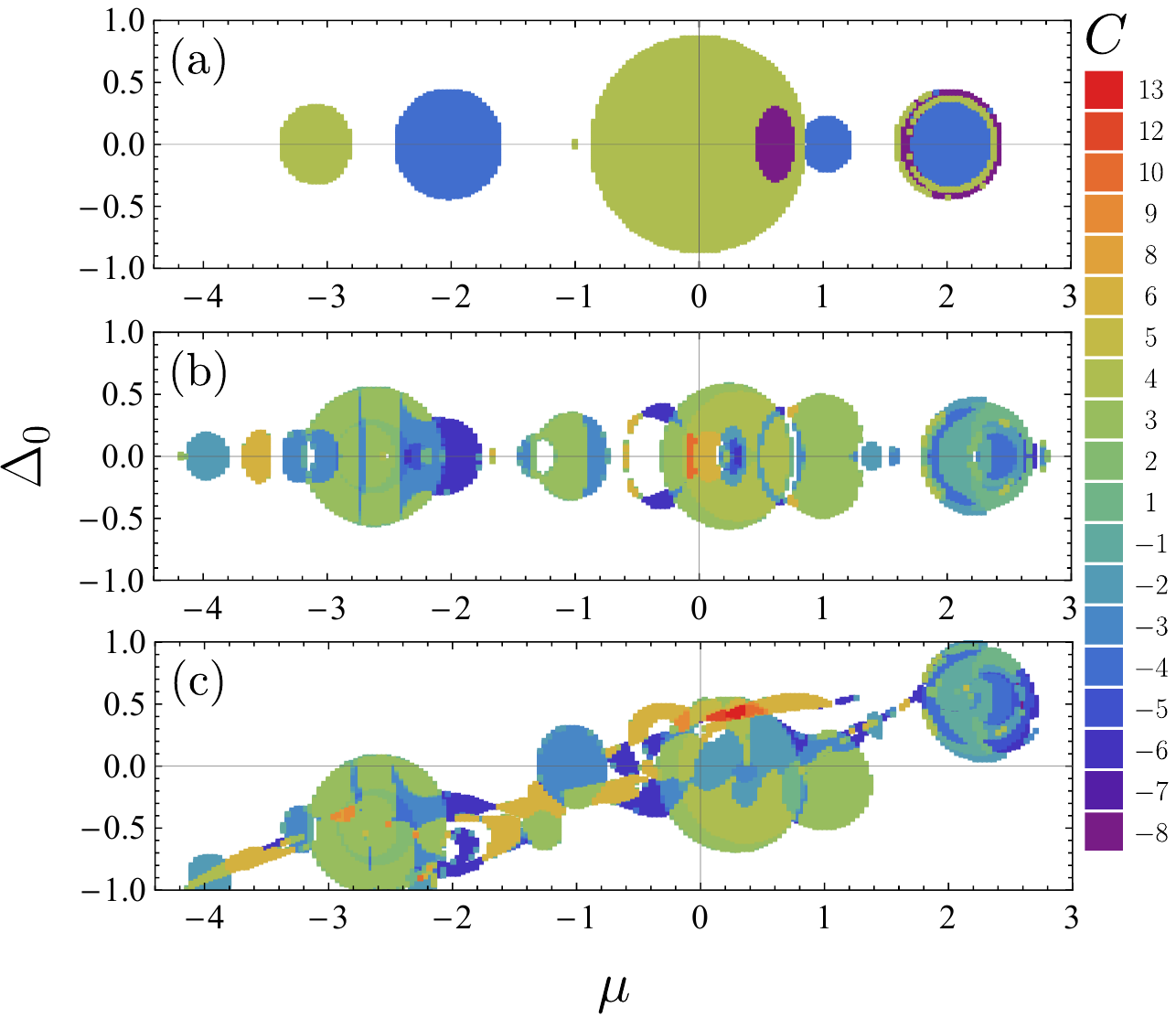}
    \caption{Topological phase diagrams of the kagome lattice with a CFP in the $(\mu,\,\Delta_0)$ space for (a) $\lambda=0$, $\Delta_1=0$, (b) $\lambda=0.5$, $\Delta_1=0$, and (c) $\lambda=0.5$, $\Delta_1=0.3$. The corresponding Chern numbers are shown in color. Odd Chern numbers are present only in panels (b) and (c). $\lambda_{\text{CFP}}=0.25$ in all panels}
    \label{FigCh}
\end{figure}

For a finite superconducting pairing, the total Chern number is calculated for the negative energy bands of the BdG Hamiltonian. 
The Chern number in a superconductor is not related to the quantization of the Hall conductance, but it determines the net number of chiral Majorana states propagating along the edge of the system.
We calculated $C$ numerically \cite{Z2Pack} in the $(\mu,\,\Delta_0)$ space as shown in Fig. \ref{FigCh}, where $\mu$ is chosen to take values along the energy spectrum of the normal state.
In the absence of RSO coupling and $\Delta_1=0$, the phase diagram reveals three (nontrivial) distinct Chern numbers $C=-8,\,-4,\,4$, as well as a trivial phase $C=0$ depending on the ($\mu$,\,$\Delta_0$) system parameters. 
The Chern number is predominantly 4 for $\mu\approx0$ and $\mu\approx-3.5$, while $C=-4$ for $\mu\approx-2$ and $\mu\approx2$ as shown in Fig.\,\ref{FigCh}(a). 
Since the presence of a MZM in the vortex core of a $\mathcal{T}$-breaking topological superconductor depends on the system's ability to support an odd number of chiral edge states \cite{Read2000}, the kagome lattice with a CFP set to $\lambda=0$ and $\Delta_1=0$ would not host MZM's.

Once RSO coupling is considered in the lattice, the spin-degenerate energy bands for $\lambda=0$ split as shown in Fig.\,\ref{CFPbands}(c), acquiring a more complicated dispersion.
Multiple Chern numbers are found as $\Delta
_0$ and $\mu$ are tuned, ranging from $C=-8$ to $C=10$ as shown in Fig.\,\ref{FigCh}(b). 
In contrast to the case with $\lambda=0$, now multiple Chern numbers around $\mu=0$ are manifest in the phase diagram, including the odd numbers $C=-5,-3,3$, and in the vicinity of $\mu=2$, large regions with Chern numbers $C=-1,1$. At lower chemical potentials, phases with $C=-3,3$ are present.

When $\Delta_1\neq0$, additional higher Chern numbers appear ($C=12,\,13$), and the phase diagram is drastically modified as shown in Fig.\,\ref{FigCh}(c), indicating a strong sensitivity of the system to nearest-neighbor electronic pairing. In the vicinity of $\mu=2$, we find large areas with odd Chern numbers $C=-5,-1,1$, while for chemical potentials around zero the Chern numbers $C=-3,1,3$ dominate. Chern numbers of $C=-3,3$ are found to dominate at lower chemical potentials.
The rich phase diagrams with even and odd Chern numbers suggest that Majorana edge states and accompanying MZM's at vortices could be achieved in these systems with properly tuned parameters.

\begin{figure*}
 \centering
    \includegraphics[scale=0.4]{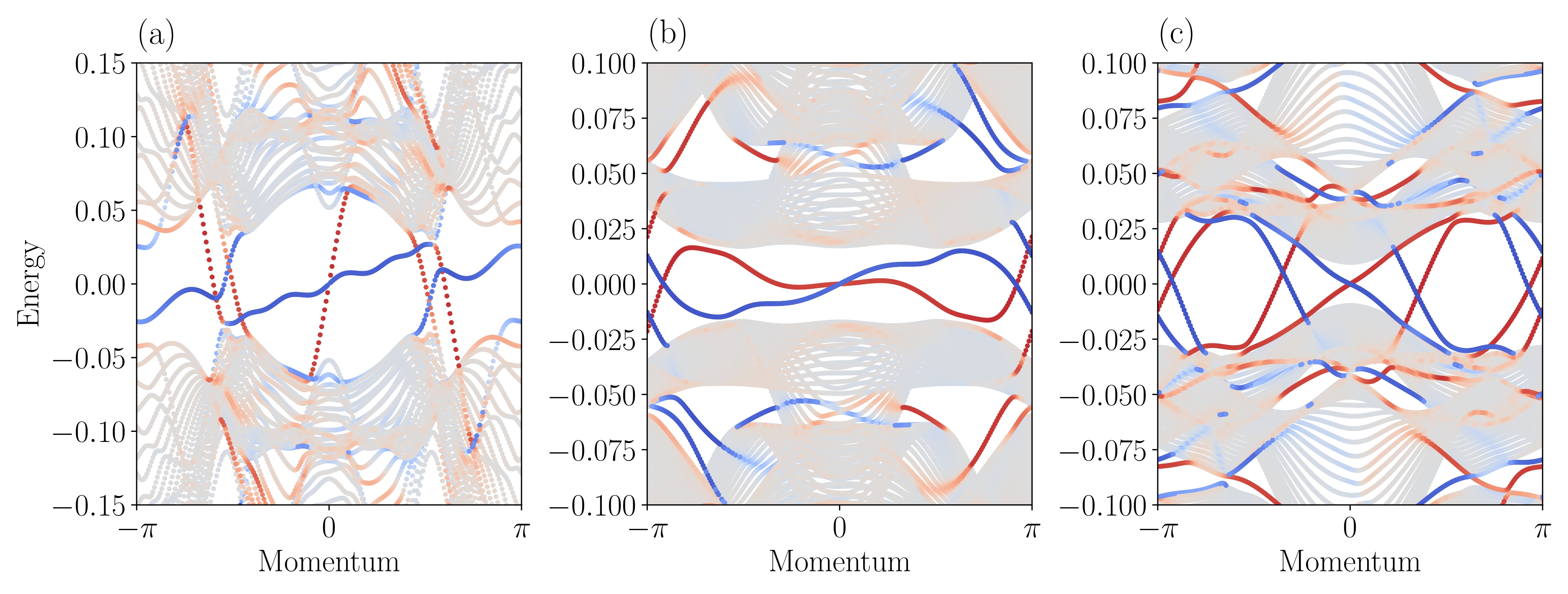}
    \caption{Energy spectrum of an infinite stripe of kagome lattice with a CFP ($\lambda_{\text{CFP}}=0.25$) considering  $\Delta_0=0.2$ and $\Delta_1=0$, with (a) $\mu=-0.2$ and (b) $\mu=2.15$, respectively. In (c) we take  $\Delta_0=0.2$ and $\Delta_1=0.3$ and $\mu=2.5$. We observe that the net number of chiral Majorana edge states shown in blue and red for the top and bottom of the strip, respectively, is equal to the bulk Chern numbers $C=3$, $C=-1$, and $C=-5$ for (a), (b) and (c), respectively. }
    \label{MES}
\end{figure*}

Fig.\,\ref{MES} shows the energy bands of an infinite strip of superconducting kagome lattice with a CFP and RSO coupling for several values of system parameters. 
When $\Delta_1=0$ and $\Delta_0=0.2$ are fixed, raising the chemical potential from $\mu=-0.2$ (Fig.\,\ref{MES}(a)) to $\mu=2.5$ (Fig.\,\ref{MES}(b)) leads to a topological phase transition from $C=-1$ to $C=3$, passing through several topological phases with different Chern numbers as shown in Fig.\,\ref{FigCh}(b).
We also show the energy bands for a strip with $C=-5$ at finite $\Delta_1=0.3$ for $\mu=2.5$ in Fig.\,\ref{MES}(c).
We observe that the Chern number does not necessarily yield the number of edge states in the energy spectrum: in the case in Fig.\,\ref{MES}(a) with $C=3$, for example, the edge states in the bottom of the strip exceed 3 (red).
However, notice that the net chirality is indeed 3, and is always preserved in all cases \cite{Zeng2023}.
In all of these cases in the vicinity of zero energy, the edge states can be described by a linear-in-momentum Hamiltonian.
A MZM is expected at vortex cores, which remains protected due to the particle-hole symmetry of the BdG spectrum.

Transitions between phases with distinct Chern numbers could be achieved by strain tuning of the superconducting pairing potential, as superconductivity \cite{Qian2021,Yin2021S} and topological phases \cite{Mojarro2023,Mojarro2024} have been shown to be highly tunable by external deformations.


\section{conclusions}\label{sec5}

We have studied the topological properties of the kagome lattice with on-site and nearest-neighbor $s$-wave superconducting pairing in the presence of RSO coupling.
When the RSO coupling strength $\lambda$ is zero, a nodal superconducting phase manifests when $\Delta_0/\Delta_1=\mu$, and it is a gapped phase otherwise.
A finite $\lambda$ leads to the breaking of spin degeneracy and corresponding energy band splitting, opening the possibility of $\mathcal{T}$-invariant topological superconducting phases for finite pairing \cite{Zhang2013}.  
By analyzing the sign of the pairing in Fermi contours we calculate the $\mathbb{Z}_2$ topological invariant \cite{Qi2010}, and identify topological, trivial, and nodal phases as the chemical potential is tuned.
We find that finite nearest-neighbor pairing is necessary to achieve topological states in the phase diagram.
In such phases, a pair of counterpropagating Majorana conducting states appears along the edge of the lattice, and are related to each other by $\mathcal{T}$. 
A Kramers pair of Majorana bound states would reside in a vortex core \cite{Qi2009}, and although it has been theoretically shown that Majorana Kramers pairs could not be used for braiding operations \cite{Konrad2016}, applying an in-plane magnetic field could result in corner-localized MZM \cite{Pahomi2020,Zhang2020}.

We have also introduced a $\mathcal{T}$-breaking chiral flux phase (CFP) in our model to investigate the possibility of chiral Majorana edge states in the absence of external magnetic fields or complex pairing symmetries. 
This CFP has been proposed to be the leading charge ordering instability in the kagome superconductors $A$V$_3$Sb$_5$, thus its experimental relevance. 
The CFP enlarges the unit cell in real space by 4 resulting in Brillouin zone folding, and the energy bands acquiring nontrivial Chern numbers \cite{Xilin2021}.
Finite on-site pairing opens a superconducting gap and a total Chern number for negative Bogoliubov bands can be calculated as a function of the chemical potential. 
This Chern number determines the \MA{net} number of chiral Majorana conducting states on the edges of the system, and generally leads to a MZM in a vortex core when the Chern number is odd \cite{Read2000}.
We find that in the absence of RSO coupling, the system acquires only even Chern numbers for any value of the chemical potential within the spectrum of the system.
However, for $\lambda\neq0$, regions with both even and odd Chern numbers appear in the phase diagrams for vanishing and non-zero $\Delta_1$. 

Our work demonstrates that Majorana edge states could be realized in kagome superconductors hosting $s$-wave superconducting pairing potential with RSO interaction. We hope these results motivate the exploration of Majorana physics in both recently discovered vanadium-based kagome superconductors as well as in other kagome materials. 




\section{acknowledgements}

We acknowledge useful discussions with Mahmoud Asmar. Supported by U.S. Department of Energy, Office of Basic Energy Sciences, Materials Science and Engineering
Division.

{\onecolumngrid
\appendix{}

\section{$2\times2$ chiral flux phase modulation}\label{Append-A}
Eqs.\,\eqref{HCFP}-\eqref{LCFP} define the CFP term in the Hamiltonian, which can be written in position space (omitting the spin)
\begin{equation}
    \hat{\mathcal{H}}_{\text{CFP}}=i\lambda_{\text{CFP}}\sum_{{\bf r},\gamma=\pm}\gamma\cos({\bf Q}_1\cdot{\bf r})\,\hat{c}^\dagger_{\text{A}{\bf r}}\hat{c}_{\text{B}{\bf r}+\gamma\bm{\delta}_1}+\gamma\cos({\bf Q}_2\cdot{\bf r})\,\hat{c}^\dagger_{\text{A}{\bf r}}\hat{c}_{\text{C}{\bf r}+\gamma\bm{\delta}_2}+\gamma\cos({\bf Q}_3\cdot{\bf r})\,\hat{c}^\dagger_{\text{B}{\bf r}+\bm{\delta}_1}\hat{c}_{\text{C}{\bf r}+\bm{\delta}_1+\gamma\bm{\delta}_3}+\text{H.c.}\,,
\end{equation}
where $\bf r$ runs over A sites.
Considering the Fourier transform of the fermionic operators $c_{\bf r}=(1/\sqrt{N})\sum_{\bf k}e^{i{\bf k}\cdot{\bf r}}c_{\bf k}$, where $N$ is the number of sites of the corresponding sublattice, we obtain
\begin{eqnarray}
\nonumber\hat{\mathcal{H}}_{\text{CFP}}&=&i\frac{\lambda_{\text{CFP}}}{N}\sum_{{\bf k}{\bf k}'}\sum_{{\bf r},\gamma=\pm}\gamma\cos({\bf Q}_1\cdot{\bf r})\,e^{-i{\bf k}\cdot{\bf r}}e^{i{\bf k}'\cdot({\bf r}+\gamma\bm{\delta}_1)}e^{i{\bf k}'\cdot\bm{\delta}_1}\hat{c}^\dagger_{\text{A}{\bf k}}\hat{c}_{\text{B}{\bf k}'} +\gamma\cos({\bf Q}_2\cdot{\bf r})\,e^{-i{\bf k}\cdot{\bf r}}e^{i{\bf k}'\cdot({\bf r}+\gamma\bm{\delta}_2)}e^{i{\bf k}'\cdot\bm{\delta}_2}\hat{c}^\dagger_{\text{A}{\bf k}}\hat{c}_{\text{C}{\bf k}'}\\
    &&\qquad\qquad\qquad\quad +\,\gamma\cos({\bf Q}_3\cdot{\bf r})\,e^{-i{\bf k}\cdot({\bf r}+\bm{\delta}_1)}e^{i{\bf k}'\cdot({\bf r}+\bm{\delta}_1+\gamma\bm{\delta}_3)}e^{-i{\bf k}\cdot\bm{\delta}_1}e^{i{\bf k}'\cdot\bm{\delta}_2}\hat{c}^\dagger_{\text{B}{\bf k}}\hat{c}_{\text{C}{\bf k}'}+\text{H.c.}\,,
\end{eqnarray}
where we have considered the gauge transformations $c_{\text{B}{\bf k}}\rightarrow e^{i{\bf k}\cdot\bm{\delta}_{1}}c_{\text{B}{\bf k}}$ and $c_{\text{C}{\bf k}}\rightarrow e^{i{\bf k}\cdot\bm{\delta}_{2}}c_{\text{C}{\bf k}}$ in order to maintain a proper periodic Bloch form $H({\bf k}+{\bf G})=H({\bf k})$, with ${\bf G}$ a reciprocal basis vector. Then 
\begin{eqnarray}
    \nonumber\hat{\mathcal{H}}_{\text{CFP}}&=&i\frac{\lambda_{\text{CFP}}}{2N}\sum_{{\bf k}{\bf k}'}\sum_{{\bf r},\gamma=\pm}\gamma\left(e^{i{\bf r}\cdot({\bf Q}_1-{\bf k}+{\bf k'})}+e^{i{\bf r}\cdot(-{\bf Q}_1-{\bf k}+{\bf k'})}\right)\,e^{i{\bf k}'\cdot\bm{\delta}_1(\gamma+1)}\hat{c}^\dagger_{\text{A}{\bf k}}\hat{c}_{\text{B}{\bf k}'}\\
    \nonumber&&\qquad\qquad\qquad\quad+\,\gamma\left(e^{i{\bf r}\cdot({\bf Q}_2-{\bf k}+{\bf k'})}+e^{i{\bf r}\cdot(-{\bf Q}_2-{\bf k}+{\bf k'})}\right)\,e^{i{\bf k}'\cdot\bm{\delta}_2(\gamma+1)}\hat{c}^\dagger_{\text{A}{\bf k}}\hat{c}_{\text{C}{\bf k}'}\\
        &&\qquad\qquad\qquad\quad+\,\gamma\left(e^{i{\bf r}\cdot({\bf Q}_3-{\bf k}+{\bf k'})}+e^{i{\bf r}\cdot(-{\bf Q}_3-{\bf k}+{\bf k'})}\right)\,e^{-i2{\bf k}\cdot\bm{\delta}_1}e^{i{\bf k}'\cdot(\bm{\delta}_1+\bm{\delta}_2+\gamma\bm{\delta}_3)}\hat{c}^\dagger_{\text{B}{\bf k}}\hat{c}_{\text{C}{\bf k}'}+\text{H.c.}\,,
\end{eqnarray}
and considering the sum representation of the delta $\delta_{{\bf k}{\bf k}'}=(1/N)\sum_{\bf r}e^{i{\bf r}\cdot({\bf k}-{\bf k}')}$, we recover
\begin{eqnarray}        \nonumber\hat{\mathcal{H}}_{\text{CFP}}&=&i\frac{\lambda_{\text{CFP}}}{2}\sum_{{\bf k}{\bf k}',\gamma=\pm}\gamma\left(\delta_{{\bf k}-{\bf Q}_1,{\bf k'}}+\delta_{{\bf k}+{\bf Q}_1,{\bf k'}}\right)\,e^{i{\bf k}'\cdot\bm{\delta}_1(\gamma+1)}\hat{c}^\dagger_{\text{A}{\bf k}}\hat{c}_{\text{B}{\bf k}'}+\gamma\left(\delta_{{\bf k}-{\bf Q}_2,{\bf k'}}+\delta_{{\bf k}+{\bf Q}_2,{\bf k'}}\right)\,e^{i{\bf k}'\cdot\bm{\delta}_2(\gamma+1)}\hat{c}^\dagger_{\text{A}{\bf k}}\hat{c}_{\text{C}{\bf k}'}\\
        \nonumber&&\quad\qquad\qquad\quad+\,\gamma\left(\delta_{{\bf k}-{\bf Q}_3,{\bf k'}}+\delta_{{\bf k}+{\bf Q}_3,{\bf k'}}\right)\,e^{-i2{\bf k}\cdot\bm{\delta}_1}e^{i{\bf k}'\cdot(\bm{\delta}_1+\bm{\delta}_2+\gamma\bm{\delta}_3)}\hat{c}^\dagger_{\text{B}{\bf k}}\hat{c}_{\text{C}{\bf k}'}+\text{H.c.}\\
       \nonumber &=&i\frac{\lambda_{\text{CFP}}}{2}\sum_{{\bf k},\gamma=\pm}\gamma\left(e^{i({{\bf k}-{\bf Q}_1})\cdot\bm{\delta}_1(\gamma+1)}+e^{i({{\bf k}+{\bf Q}_1})\cdot\bm{\delta}_1(\gamma+1)}\right)\,\hat{c}^\dagger_{\text{A}{\bf k}}\hat{c}_{\text{B}{\bf k}+{\bf Q}_1}\\
       \nonumber&&\qquad\qquad\quad+\,\gamma\left(e^{i({{\bf k}-{\bf Q}_2})\cdot\bm{\delta}_2(\gamma+1)}+e^{i({{\bf k}+{\bf Q}_2})\cdot\bm{\delta}_2(\gamma+1)}\right)\,\hat{c}^\dagger_{\text{A}{\bf k}}\hat{c}_{\text{C}{\bf k}+{\bf Q}_2}\\
       \nonumber &&\qquad\qquad\quad+\,\gamma\left(e^{i({{\bf k}-{\bf Q}_3})\cdot(\bm{\delta}_1+\bm{\delta}_2+\gamma\bm{\delta}_3)}+e^{i({{\bf k}+{\bf Q}_3})\cdot(\bm{\delta}_1+\bm{\delta}_2+\gamma\bm{\delta}_3)}\right)\,e^{-i2{\bf k}\cdot\bm{\delta}_1}\hat{c}^\dagger_{\text{B}{\bf k}}\hat{c}_{\text{C}{\bf k}+{\bf Q}_3}+\text{H.c.}\\
        &=&i\lambda_{\text{CFP}}\sum_{{\bf k},\gamma=\pm}\gamma e^{i{\bf k}\cdot\bm{\delta}_1(\gamma+1)}\,\hat{c}^\dagger_{\text{A}{\bf k}}\hat{c}_{\text{B}{\bf k}+{\bf Q}_1}+\gamma e^{i{\bf k}\cdot\bm{\delta}_2(\gamma+1)}\,\hat{c}^\dagger_{\text{A}{\bf k}}\hat{c}_{\text{C}{\bf k}+{\bf Q}_2}-\gamma e^{i{\bf k}\cdot\bm{\delta}_3(\gamma+1)}\,\hat{c}^\dagger_{\text{B}{\bf k}}\hat{c}_{\text{C}{\bf k}+{\bf Q}_3}+\text{H.c.}\,.
\end{eqnarray}
In the last line we have taken into account ${\bf Q}_i\cdot
\bm{\delta}_i=0$, ${\bf Q}_3\cdot
\bm{\delta}_1=\pi/2$, and $\bm{\delta}_3=\bm{\delta}_2-\bm{\delta}_1$. We have also considered that $\hat{c}_{{\bf k}-{\bf Q}_i}=\hat{c}_{{\bf k}+{\bf Q}_i}$, due to the periodicity of the reciprocal lattice. Finally
\begin{equation}
    \hat{\mathcal{H}}_{\text{CFP}}=i\lambda_{\text{CFP}}\sum_{{\bf k}}f_-(k_1)\,\hat{c}^\dagger_{\text{A}{\bf k}}\hat{c}_{\text{B}{\bf k}+{\bf Q}_1}+f_-(k_2)\,\hat{c}^\dagger_{\text{A}{\bf k}}\hat{c}_{\text{C}{\bf k}+{\bf Q}_2}-f_-(k_3)\,\hat{c}^\dagger_{\text{B}{\bf k}}\hat{c}_{\text{C}{\bf k}+{\bf Q}_3}+\text{H.c.}\,,
\end{equation}}
where $k_i=\textbf{k}\cdot\bm{\delta}_i$ and $f_{\pm}(x)=e^{2ix}\pm1$.

\twocolumngrid

\bibliography{biblio.bib}

\begin{thebibliography}{58}
\bibitem [{not()}]{note1}%
  \BibitemOpen
  \href@noop {} {\bibinfo {title} {We take $\lambda=0.5$ throughout this wor. although smaller values are likely, they would result in qualitatively similar physics, if with an adjusted scale.}}\BibitemShut {Stop}%
\end{thebibliography}
\end{document}